\documentstyle{mn}
\input{psfig.sty}

\begin{document}

\title{Does magnetic pressure affect the ICM dynamics?}
\author[D. R. Gon\c calves and A. C. S. Fria\c ca]
        {D. R. Gon\c calves$^{1,2}$ and A. C. S. 
         Fria\c ca$^{1}$ \\
$^{1}$ Instituto Astron\^omico e Geof\'\i sico - USP, Av. Miguel Stefano 4200,
04301-904 S\~ao Paulo, SP, Brazil \\
$^{2}$ Instituto de Astrof\'\i sica de Canarias, E-38200 La Laguna, Tenerife, 
Spain}
\date{denise@ll.iac.es, amancio@iagusp.usp.br}
\maketitle

\begin{abstract}
A possible discrepancy found in the determination of mass from gravitational
lensing data, and from X-rays observations, has been largely discussed in
the latest years. For instance, Miralda-Escud\'{e} \& Babul (1995) have
found that the mass estimate derived from gravitational lensing can be as
much as a factor of $2-2.5$ larger than the mass estimate derived from
analysis of the X-rays observations. Another important discrepancy related
to these data is that X-ray imaging, with some spectral resolution, suggest
that the mass distribution of the gravitating matter, mostly dark matter,
has a central cusp, or at least that the dark matter is more centrally
condensed than the X-ray-emitting gas, and also with respect to the galaxy
distribution (Eyles et al. 1991), at variance to what is expected from the most
accepted models of formation of large scale structure. Could these
discrepancies be consequence of the standard description of the ICM, in
which it is assumed hydrostatic equilibrium maintained by thermal pressure?
In analogy to the interstellar medium of the Galaxy, it is expected a
non-thermal term of pressure, which contains contributions of magnetic
fields, turbulence and cosmic rays. We follow the evolution of the ICM,
considering a term of magnetic pressure, aiming at answering the question
whether or not these discrepancies can be explained via non-thermal terms of
pressure. Our results suggest that the magnetic pressure could only affect
the dynamics of the ICM on scales as small as $\la 1 \;{\rm kpc}$.  
Our models are constrained by the observations of large and small scale fields 
and we are successful at reproducing available data, for both Faraday
rotation limits and inverse Compton limits for the magnetic fields. 
In our calculations the radius (from the cluster center) in which magnetic
pressure reaches equipartition is smaller than radii derived in previous 
works. The crucial difference in our models comes from our more
realistic treatment of the magnetic field geometry, and from the
consideration of a sink term in the cooling flow which reduces the
amplification of the field strength during the inflow. In addition the 
magnetic field calculations are changed after the cooling flow has been formed.
\end{abstract}


\section{Introduction}

Since the work of Loeb \& Mao (1994), the possibility of
explaining the discrepancies on mass determinations, found by
Miralda-Escud\'{e} \& Babul (1995), via non-thermal pressure support
 has been widely discussed (see also Wu \& Fang 1996, 1997; Wu et al. 1998). 
 The discrepancy 
arises from the two most promising techniques to obtain clusters of
galaxies masses. On one hand, the determination of masses in clusters of
galaxies, via X-ray data, is based on the hypothesis that the ICM is in
hydrostatic equilibrium with the gravitational potential, using the radial
profiles of density and temperature. There are uncertainties in the
determination of temperature profiles, particularly for radii $>1\;{\rm Mpc}$,
and for most systems only a mean emission-weighted X-ray temperature is
available (radial temperature profiles are available only for a few
clusters ,e.g., Allen \& Fabian 1994; Nulsen \& B\"{o}hringer 1995). On the
other hand, gravitational lensing measures the projected surface density of
matter, a method which makes no assumptions on the dynamical state
of the gravitating matter (Fort \& Mellier 1994; Miralda-Escud\'{e} \& Babul
1995; Smail et al. 1997).

One can find in the literature some attempts to resolve the discrepancy
between X-ray and gravitational lensing mass measurements of clusters of
galaxies. For instance, Allen (1998) studied in detail a sample of 13 galaxy
clusters (including cooling flows,
intermediate and non-cooling flows systems) with the goal of comparing X-ray
and lensing mass measurements. His conclusions pointed out that, at least
for cooling flows systems, being more relaxed systems, this discrepancy is
completed resolved, and therefore, non-thermal pressures can be discarded in
these systems.

The magnetic field of the ICM can be obtained via Faraday rotation, due to
the effect of magnetic field on the polarized radio emission from the
cluster or the background radio sources. The polarization plane of linearly
polarized radiation is rotated during the passage through a magnetized
plasma. The angle of rotation is $\phi =(RM)\lambda ^2$, where $RM$ is the
rotation measure and $\lambda $ the radiation wavelength (Sarazin 1992, for
a review). In clusters with diffuse radio emission, X-ray observations can
give a lower limit to the strength of the magnetic field. Typically, this
limit is $B\geq 0.1\;{\rm \mu G}$ (Rephaeli et al. 1987) on scales of $\sim
1\;{\rm Mpc}$. In the case of Faraday rotation the information obtained is the upper
limit on the intensity of the field, and the measured values are $RM\leq
100\;{\rm rad/m^2}$, that is more or less consistent with a intracluster
field of $B\sim 1\;{\rm \mu G}$, with a coherence length of $l_B\leq 10\;{\rm kpc%
}$. This strength of the magnetic field corresponds to a ratio of magnetic
to gas pressure of $p_B/p_{gas}$ $\leq 10^{-3}$, implying that $B$ does not
influence the cluster dynamics (at least on large scales).

At inner regions, of the cooling flow clusters,
 the magnetic fields are expected to be amplified due to the
gas compression (Soker \& Sarazin 1990). If they are frozen in the mass flow flux, and
if this flux is homogeneous and spherically symmetric, $B\propto r^{-1}$ and $%
RM\propto r^{-1}$, ($p_B\propto r^{-2}$ and the gas pressure increases
slowly). Even in this case $p_B$ reaches equipartition at a radius $r_B$ of $%
r_B\sim 1\;{\rm kpc}\left( \frac B{1\;{\rm \mu G}}\right) ^{1/2}\left( {\dot{M}%
}/100\;{\rm M}_{\odot }\;{\rm yr ^{-1}}\right) ^{1/3}$.\ In these inner regions many
sources with very strong Faraday rotations were observed, in which the
rotation measure can reach values of $RM\sim 4000\;{\rm rad/m^2}$ (radio
sources associated with the central galaxies of the clusters with very
strong cooling flows (M87/Virgo, Cyg A, Hydra A, 3C 295, A1795)), implying, $%
B\geq 10\;{\rm \mu G} $ at $l_B\sim 1\;{\rm kpc}$ (Taylor \& Perley 1993; Ge
\& Owen 1993, 1994). These observations strongly suggest that the Faraday
rotation is created by magnetic fields within the cooling flow clusters.

Another promising method to estimate the cluster scale magnetic field, as 
cited above, is the
detection of co-spatial inverse Compton X-ray emission with the synchrotron
plasma emission (the $3\;{\rm K}${\ background photons scattering off the
relativistic electrons can produce a diffuse X-ray emission) (Rephaeli \&
Gruber 1988). Therefore, this method provides limits on the cluster scale
magnetic fields, in addition of limits on the non-thermal amount of X-ray
emission (or even on the relativist electrons energy) in galaxy clusters.
Such a kind of detection of clusters magnetic fields leads, using ROSAT\
PSPC data and also }${327}\;{\rm MHz}$ radio map of Abell 85 (a cooling flow
cluster, with a central dominant cD galaxy and about $100\;{\rm M}_{\odot}
{\rm /yr}$), to an estimate of $(0.95\pm 0.10)\;{\rm \mu G}$
(Bagchi et al. 1998). However, even non-cooling flows clusters present this
diffuse, relic radio source which can be used to estimate magnetic field
strength. For instance Ensslin \& Biermann (1998) studied limits on the Coma
cluster magnetic field strength, using this multifrequency observations.
They shown that the central magnetic field limit is {{$B>0.3\;{\rm \mu G}$.
Others have determined the strength of the magnetic field for Coma cluster,
using different techniques and obtaining similar values: $B\leq 1.2\;{\rm %
\mu G}$ (Lieu et al. 1996); {$B>0.4\;{\rm \mu G}$} (Sreekumar et al. 1996).
For the same cluster (Coma), but using Faraday rotation measure, Feretti et
al. (1995) estimated magnetic fields of }}$6.0${$\;{\rm \mu G}$ (at scales
of }$1\;{\rm kpc}$), and of {{$1.7\;{\rm \mu G}$ (at scales of $10\;{\rm kpc}$)
was estimated by Kim et al. (1990). }}

The above scenario allow us conclude that for both methods the observational
resolution of the telescope limits the detection of smaller scales magnetic
fields, implying that at scales smaller than $1\;{\rm kpc}$ the magnetic field
strength can be higher (Ensslin et al. 1997). Another point to be noted is
that Faraday rotation measures always gives values higher than inverse
Compton/CBM measures. Anyway, these fields are present in the ICM and
therefore justify the such a kind of study we present here. Other
theoretical works concerning the magnetic pressure on the ICM are available
(for instance Soker \& Sarazin 1990; Tribble 1993; Zoabi et al. 1996) and we
briefly compare our results with those obtained by these authors.

Our goal in this paper is trying to answer the question whether or not
magnetic support can be relevant in cooling flow clusters, using a more
realistic treatment of the magnetic field geometric evolution. The scope of
the paper is the following: in Section 2 we present the hydrodynamical 
equations and the method applied for their solution; Section 3 describes 
our models and results compared to the available observations; and in 
Section 4 we discuss our results in the light of others obtained in 
previous works, as well our main conclusions.

\section{Evolution of the ICM with Magnetic Pressure}

The evolution of the intracluster gas is obtained by solving the
hydrodynamic equations of mass, momentum and energy conservation:

\begin{equation}
\frac{\partial \rho }{\partial t}+\frac 1{r^2}\frac \partial {\partial
r}\left( r^2\rho u\right) =-\omega \rho
\end{equation}

\begin{equation}
\frac{\partial u}{\partial t}+u\frac{\partial u}{\partial r}=-\frac 1\rho 
\frac{\partial p_t}{\partial r}-\frac{GM(r)}{r^2}
\end{equation}

\begin{equation}
\frac{\partial U}{\partial t}+u\frac{\partial U}{\partial r}=\frac{p_t}{\rho
^2} \left( \frac{\partial \rho}{\partial t}+u\frac{\partial \rho }{\partial r}%
\right) -\Lambda \rho
\end{equation}

\noindent where $u$, $\rho $, $p_t$, $U$ are the gas velocity, density,
total pressure and the specific internal energy. The equation of state
relates $U$ and the temperature,

\begin{equation}
U=\frac 32\frac{k_BT}{\mu m_H}
\end{equation}

\noindent  ($k_B$ is the Boltzmann's constant, $m_H$ is the hydrogen atom
mass and $\mu =0.62$ is the mean molecular weight of a fully ionized gas
with $10\%$ helium by number). The mass distribution, $M(r)$, is due to the
contribution of the X-rays emitting gas plus the cluster collisionless matter
(which is the sum of the contributions of galaxies and dark matter -- the
latter being dominant), i.e, $M(r)=M_g(r)+M_{cl}(r)$. $M_{cl}(r)$ follows

\begin{equation}
\rho_{cl} (r)=\rho_0 \left( 1+\frac{r^2}{a^2} \right)^{-3/2}
\end{equation}

\noindent  in which {$\rho _0$ and $a$ (the cluster core radius) are related
to $\sigma $ (the line-of-sight velocity dispersion) via: 
$9\sigma ^2=4\pi Ga^2\rho _0$. }

The total pressure $p_t$ is the sum of thermal and magnetic pressure, e.g., 
$p_t=p+p_B$. The constraints to the magnetic pressure come from observations,
from which $p_B=B^2/8\pi \simeq 4\times 10^{-14}\;{\rm erg\;cm}^{-3} \;{\rm s}^{-1}$
(cf. Bagchi et al. 1998) for a diffuse field located at $\sim 700h_{50}^{-1}%
\;{\rm kpc}$ from the cluster center. Along this paper we will use mostly the
ratio between magnetic and thermal pressures, or the $\beta $-parameter, $%
\beta =p_B/p$.

The sink term $\omega \rho $ in the mass equation describes the removal of mass
from the gas flow by thermal instabilities. The importance of the gas removal 
was studied in detail by Fria\c {c}a (1993) following
the q-description described by White \& Sarazin (1987). The sink is particularly
important when one searches for a steady state solution of the cooling flow
without an implausible huge accumulation of mass at the center. In fact, the
condensations formed by the sink will probably give rise to stars, planetary
bodies or cold dense clouds which in turn will constitute a halo surrounding
the central dominant galaxy. We assume isobaric removal, so that the sink
does not introduce any additional term in the energy equation. Summing up
the physics contained in this term one can say that the specific mass
removal rate is $\omega =q/t_c$ where the denominator is the instantaneous
isobaric cooling time, such that the removal efficiency $q$ relates the
cooling time to the growth time scale of the thermal instability in the
cooling flow. We assume $q$ between $1.0$ and $1.5$, which are the $q$-values 
found to be more consistent with the observations (Fria\c {c}a 1993).

The cooling function adopted $\Lambda (T)$ is the cooling rate per unit
volume. Since there is no ionization equilibrium for temperatures lower than 
$10^6$ K, we adopt a non-equilibrium cooling function for the gas at $T<10^6$ K 
(the recombination time of important ions is longer than the cooling time at
these temperatures). The cooling function was calculated with the atomic
database of the photoionization code AANGABA (Gruenwald \& Viegas 1992). The
adopted abundances are sub-solar as appropriate for the ICM (Edge \& 
Stewart 1991; Fabian 1994; Grevesse \& Anders 1989).

Despite the presence of steep temperature gradients we did not consider
thermal conduction in our models. This can be justified using the fact that
on a global scale cooling flow clusters contain cooler gas near the center
and hotter gas further out. Therefore, the presence of cooling flows is
itself a proof that thermal conduction effect is, at least, reduced in the
ICM. Models show that thermal conduction would erase the observed density
and temperature gradients in cooling flows, unless it is inhibited (see, for
instance, Fria\c {c}a 1986; David \& Bregman 1989). It is well known that
even weak magnetic field, if it is tangled, can inhibit the thermal
conduction perpendicular to the field lines. More recently it has been
argued that electromagnetic instabilities driven by temperature gradients
(or electric currents in other situations) also can cause this inhibition in
cooling flows (Pistinner et al. 1996), even for non-tangled field lines. 

A spherically symmetric Eulerian code is employed for the calculations, which
are solved via the finite-difference scheme based on Cloutman (1980). The
grid points are spaced logarithmically, with a grid of 100 cells, with the
first being $50\;{\rm pc}$ wide. The innermost cell edge is located at 
$100\;{\rm pc}$ and the outer boundary at twice the tidal radius of the
cluster. The artificial viscosity for the treatment of the shocks follows
the formulation of Tscharnuter \& Winkler (1979) based on the Navier-Stokes
equation. The outer boundary conditions on pressure and density are derived
by including an outer fictitious cell, the density and pressure in which are
obtained from extrapolation of power laws over the radius fitted to the five
outermost real cells. The inner boundary conditions are adjusted according
to whether inflow (velocity at the inner boundary is extrapolated from the
velocities at the innermost cell edges) or outflow (velocity is set zero)
prevails locally. The initial conditions for the gas are an isothermal
atmosphere ($T_0=10^7\;{\rm K}$) with $30\%$ solar abundances and
density distribution following that of the cluster dark matter. The
evolution is followed until the age of $14\;{\rm Gyr}$.

The initial $\beta $ value used here was derived from the magnetic field
observations (using, for instance, Bagchi et al. 1998; Ge \& Owen 1993,
1994; Ensslin \& Biermann 1998; Ruy \& Biermann 1998). We assume: frozen-in
field; spherical symmetry for the flow and the cluster itself; and that at $r
> r_c$ (the cooling radius, see below), the magnetic field is isotropic, i.e.,

\[
B_{r}^2=B_{t}^2/2=B^2/3 
\]

\noindent and $l_{r}=l_{t}\equiv l$ (where $B_{r}$ and $B_{t}$ are the
radial and transversal components of the magnetic field $B$ and $l_{r}$ and $%
l_{t}$ are the coherence length of the large-scale field in the radial and
transverse directions). In order to calculate $B_{r}$ and $B_{t}$ for $r <
r_c$ we modified the calculation of the magnetic field of Soker \& Sarazin
(1990) by considering an inhomogeneous cooling flow (i.e. $\dot{M}_i\neq 
\dot{M}$ varies with $r$). Therefore, the two components of the field are
then given by

\[
{\frac{D }{Dt}}\left( B_r^2 r^{4} \dot{M}^{-2} \right) = 0 
\]

\noindent and

\[
{\frac{D }{Dt}}\left( B_t^2 r^{2} u^{2} \dot{M}^{-1} \right) = 0 . 
\]

\noindent In our models we take as reference radius the
cooling radius $r_c$. In fact we modify the geometry of the field when and
where the cooling time comes to be less than $10^{10}\;{\rm yr}$ (usually
adopted as the condition for the development of a cooling flow). Therefore,
our condition to assume a non-isotropic field is $t_{coo}\equiv 3 k_B T / 2
\mu m_H \Lambda (T) \rho \le 10^{10}\;{\rm yr}$. After the
formation of the cooling flow, in the inner regions of the ICM, the magnetic
field geometry is changed, following the enhancement of the radial component
of the field, due to the enhancement of the density. 

\section{Models and Results }

In this section we present the results of our models. There are four
parameters to consider in each one of the models: $\sigma $, the cluster
velocity dispersion; $\rho _0$, the initial average mass density of the gas; 
$a$, the cluster core radius; and $\beta _0$, the initial magnetic to
thermal pressure ratio. We adopted the removal efficiency $q=1.5$.

The most important results of our models are shown on figures we describe below,
for which we assume: $\sigma =1000\;{\rm km\;s^{-1}}$ and $a=250\;{\rm kpc}$. 
First of all, the evolution we follow here is characteristic of cooling flow
clusters and in this scenario we discuss the evolution of the basic
thermodynamics parameters. Considering the overall characteristics of our
models, we will compare the results with the very recent study based on
ROSAT observations of the cores of clusters of galaxies, by Peres et al.
(1998), focusing on cooling flows in a X-rays flux-limited sample
(containing the brightest 55 clusters over the sky in the $2-10\;{\rm keV}$
band). Comparing the present models with Peres et al. (1998) deprojection
results, we see that the central cooling time here adopted as our cooling
flow criterion, e.g. $t_{coo} \la 10^{10}\;{\rm yr}$, is typical for a fraction
between $70\%$ and $90\%$ of their sample. They also discuss briefly the
cooling flow age, remembering that in hierarchical scenarios for the
formation of structures in the Universe, clusters are formed by smaller
substructures by mergers, and therefore the estimation of the cooling flows
ages (and the cluster ages themselves) is complicated. Anyway they determine
the fraction of cooling flow clusters in their sample considering a factor
of two in the ages and concluding that the fraction do not vary that much
(from $13\;{\rm Gyr}$ to $6\;{\rm Gyr}$, the fraction varies from $70\%$ to 
$65\% $). This allow us conclude that our models, which present cooling flows
since the cluster has the age of $\sim 7-9\;{\rm Gyr}$, are typical for their
sample. As a matter of fact, the time in which the cooling flow structure is
formed depends strongly on the initial density we adopted. For models with  
$\rho _0=1.25\times 10^{-28} \;{\rm g\;cm}^{-3}$ it rises on $\sim 9\;{\rm %
Gyr}$, while the models with $\rho _0=1.5\times 10^{-28} \;{\rm g\;cm}^{-3}$
have it formed on $\sim 7\;{\rm Gyr}$. We will come back to this point later
while analyzing the field anisotropy.

The characteristics of our models are summarized using four typical set of
initial parameters, and discussing some details which came up of the study
of a larger grid of parameters. Therefore, each model is characterized by  
its position in the $(\rho _0,\beta _0)$ parameter space: model I 
$(\rho_0=1.5 \times 10^{-28} \;{\rm g\;cm^{-3}},\beta _0=10^{-2})$; 
model II $(\rho_0=1.5 \times 10^{-28} \;{\rm g\;cm^{-3}},\beta _0=10^{-3})$; 
model III $(\rho_0=1.25 \times 10^{-28} \;{\rm g\;cm^{-3}},\beta _0=10^{-2})$; and 
model IV $(\rho_0=1.25 \times 10^{-28} \;{\rm g\;cm^{-3}},\beta _0=10^{-3})$.

Figure 1 shows the evolution of density and temperature profiles
corresponding to model I, from which the presence of the cooling flow on
later stages of the ICM evolution and at inner regions is remarkable if one 
notices the steep gradients of these quantities. In
order to better understand how the magnetic field geometry is modified after the
cooling flow formation, e.g., after the steepness on the temperature and
density gradients, we follow the evolution of the degree of anisotropy,
using the concepts previously defined on Section 2, concerning the geometry
of the magnetic field. Hereafter we called `degree of anisotropy' the ratio $%
B_t/B_r$, noting that for the isotropic case it results $\sqrt{2}$ and the
more anisotropic the field geometry the smaller is this ratio. Therefore, we
present on Figure 2 the evolution of the degree of anisotropy since 
$\sim 3.3\;{\rm Gyr}$, comparing models I and III, in which one can see, 
clearly, that
the anisotropy begins decreasing on earlier times for models with higher 
$\rho _0$ (model I) than for the ones with lower values of $\rho _0$ (model
III). From Figure 2 we are allowed to conclude that the degree of anisotropy 
can be seen as a sensor of the presence of the cooling flow. In another words, 
the change in the degree of anisotropy can be used as another criterion to
indicate the epoch, on the ICM evolution, in which the cooling flow appear.

\begin{figure}
\begin{center}
\vbox{
\psfig{file=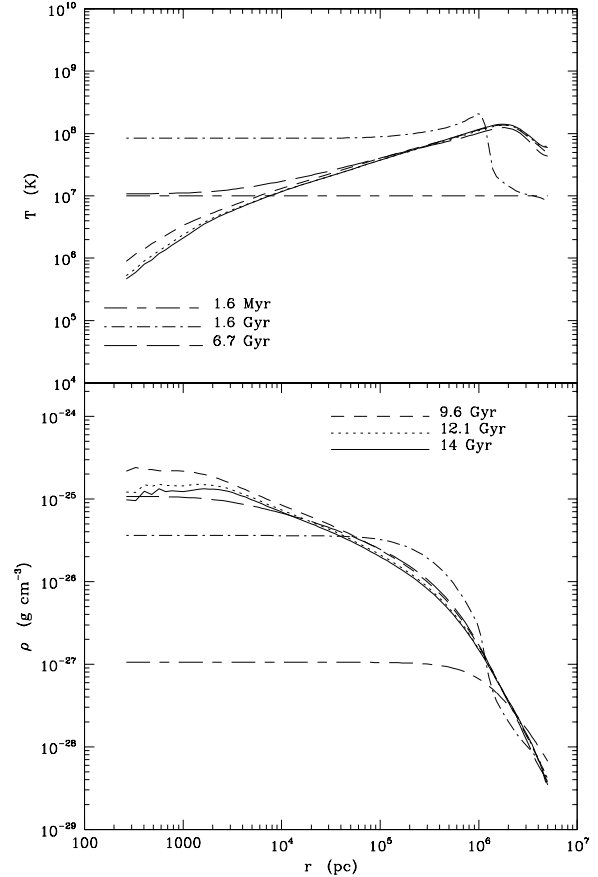,width=8.0truecm,bbllx=117pt,bblly=171pt,bburx=471pt,bbury=695pt}}
\end{center} 
\caption[]{Evolution of the density and temperature profiles. Curves represent
early and late stages of the ICM evolution, as labeled, for model I. 
Noting the steep gradients of these quantities, i.e., the presence of the
cooling flow, at inner regions of the cluster, on later stages of the ICM 
evolution.}
\end{figure}

\begin{figure}
\begin{center}
\vbox{
\psfig{file=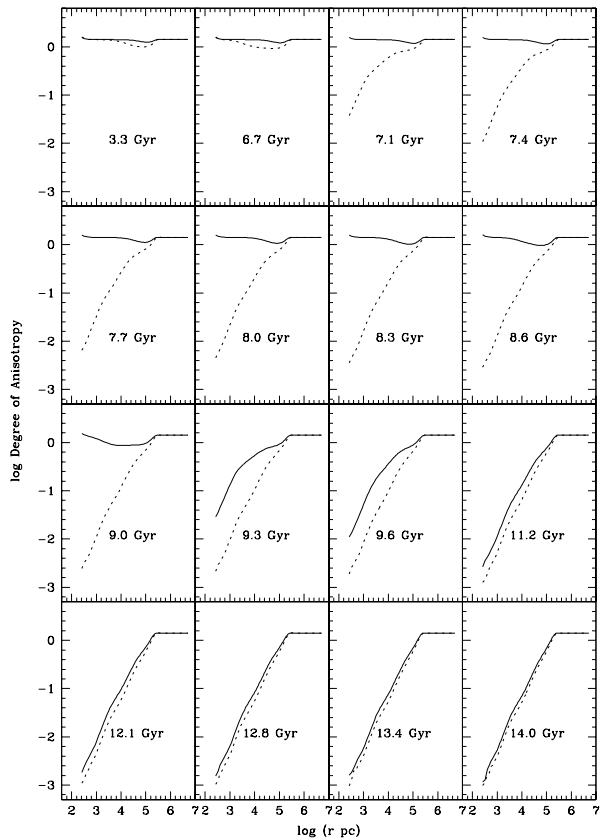,width=8.0truecm,bbllx=94pt,bblly=179pt,bburx=460pt,bbury=693pt}}
\end{center}
\caption[]{Evolution of the anisotropy degree, the tangential to radial 
magnetic field components ratio, for model I (dashed lines) and III 
(full lines), on late stages of the ICM evolution. Noting that the presence 
of the cooling flow, at inner regions, $\la 200 \;{\rm kpc}$, and on late 
stages, $\ga 9.6 \;{\rm Gyr}$, of the ICM evolution 
(see Figure 1) matches very well with the decrease of the anisotropy degree, 
and that it occurs earlier for model I.}
\end{figure}

These results can also be discussed in the light of some observational works
in which the limits to the magnetic field strength on large and small scales
of the cooling flow clusters are given. Following such a kind of
observations, as previously seen in the introduction section, we chose 
two values of
magnetic field strength derived by the authors below. The first one is
presented in Bagchi et al. (1998) who estimated, from inverse Compton X-ray
emission with the synchrotron emission plasma, a cluster-scale ($700\;{\rm kpc}
$) magnetic field strength of $(0.95\pm 0.10)\; {\rm \mu G}$ for Abell
85 (a cooling flow cluster with a central dominant cD galaxy and $\dot{M}$ $%
\simeq 100\;{\rm \ M}_{\odot }{\rm /yr}$). The second one is presented in two
papers of Ge \& Owen (1993, 1994), in which they present and discuss
rotation measures and the related intensity of the magnetic field, giving a
range of this intensity at scales of $10\;{\rm kpc}$. Therefore, our results
for the magnetic field strength and also for pressures, on large
and small scales, are compared to the chosen observed ones, in Figures 3 and
4. Reminding that the time on which the cooling flow arises is closely
related to {$\rho _0$, one can expect distinct results on the evolution of 
the field intensity from, for instance, model I to model III. However this 
evolution can be better explained comparing model I (Figure 3) to model II 
(Figure 4), since these two models have the same initial density but distinct 
$\beta_0$. }

\begin{figure}
\begin{center}
\vbox{
\psfig{file=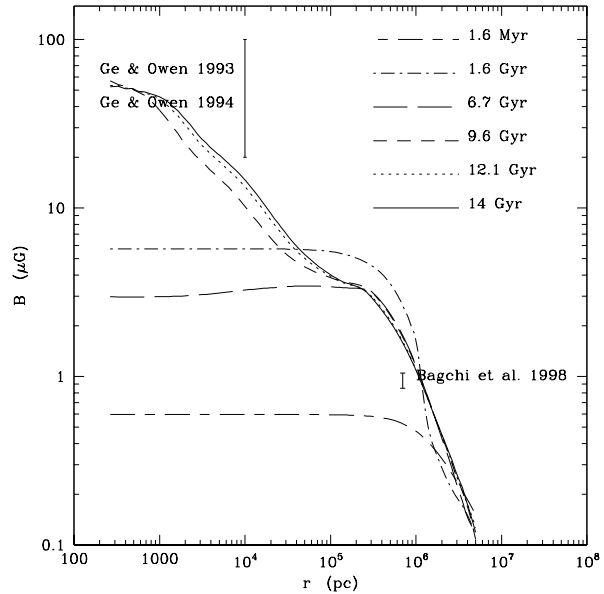,width=8.0truecm,bbllx=125pt,bblly=247pt,bburx=465pt,bbury=590pt}}
\end{center}
\caption[]{Evolution of the magnetic strength profiles compared to the 
observations. 
Curves represent early and late stages of the ICM evolution, as labeled, 
for model I ($\rho_0=1.5 \times 10^{-28} \;{\rm g\;cm^{-3}};\;
\beta_0 = 10^{-2}$). Note that the intensity $B$ of the magnetic field
at 6.7 Gyr is smaller than at 1.6 Gyr, due to the fact that at 6.7 Gyr the ICM
is in the verge of developing a cooling flow, as we can see from the drop in
temperature in the central region, as shown in Figure 1.  In the evolution of
the ICM before the onset of the cooling flow, the magnetic pressure keeps track
of the thermal pressure, following the initial conditions for $p_B/p=\beta_0 <
1$, and the reduction in the thermal pressure just after the onset of the
cooling flow is reflected in the evolution of $B$.  Only after the cooling flow
has been established, leading to amplification of $B$, the intensity of the
magnetic field will rise to high values.}
\end{figure}

\begin{figure}
\begin{center}
\vbox{
\psfig{file=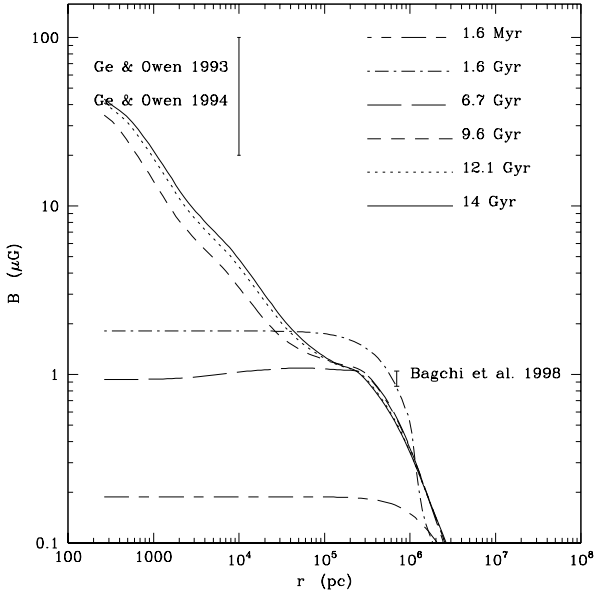,width=8.0truecm,bbllx=125pt,bblly=247pt,bburx=465pt,bbury=590pt}}
\end{center}
\caption[]{Evolution of the magnetic strength profiles compared to observations. 
Curves represent early and late stages of the ICM evolution, as labeled, 
for model II ($\rho_0=1.5 \times 10^{-28} \;{\rm g\;cm^{-3}};\;
\beta_0 = 10^{-3}$). See the comments on the evolution of $B$ made in
the caption of Figure 3.}
\end{figure}

Our best model, in terms of the magnetic field strength compared with
observations, is model I 
$(\rho _0=1.5 \times 10^{-28}\;{\rm g\;cm}^{-3},\beta _0=10^{-2})$. 
From Figure 3 it is possible to see that on scales
of $700\;{\rm kpc}$ the magnetic field expected for the model is higher than
the observed one (considering, of course, the profile correspondent to
redshift zero, or evolution times on the order of $13-14\;{\rm Gyr}$), while on scales
of $10\;{\rm kpc}$ the model gives a value lower than the observed one.
Meanwhile, at least on scales of $700\;{\rm kpc}$, the situation is inverted
if one takes a look on Figure 4, for which $\rho _0=1.5 \times 10^{-28}
\;{\rm g\;cm}^{-3}$, but $\beta _0=10^{-3}$. Given the uncertainties
characteristics of the observations, we can say that our models are in 
agreement with the magnetic field estimations available.

On Figures 5 and 6 we show the magnetic and thermal pressures evolution, or
in another words, $\beta $-evolution, for models I and II respectively, on
later times of the ICM evolution, in order to analyze when and where
magnetic pressure reaches equipartition. Obviously the magnetic pressure is
compatible with the magnetic field intensities and may be
compared to the values determined by, for instance, Bagchi et al. (1998), $%
p_B=B^2/8\pi \simeq 4\times 10^{-14}\;{\rm erg\;cm}^{-3}\;{\rm s}^{-1}$, at scales
of $700\;{\rm kpc}$, on the present time. From the analysis of the magnetic 
pressures
expected from our models it is clear that they agree, as well as the
magnetic field strength, with the observations. Here again model I appears
being the best one, with $(\beta _0=10^{-2})$, but the values expected from
model II are not far away from the observed ones as well. Noting also that magnetic
pressure and/or magnetic intensity does not change very much after $12\;{\rm %
Gyr}$, for both cases. Results presented on Figures 3 - 6 would indicate
that we should adopt an intermediate initial value for $\beta$ (like 
$\beta_0=5\times 10^{-3}$) in order to obtain a magnetic field intensity in better
agreement with the observations, at least on scales of $700\;{\rm kpc}$.
Nevertheless such an exercise should not solve the match of models and
observations on smaller scales, since  $\beta _0$ 
$\simeq 5\times 10^{-3}$ should decrease magnetic pressure on scales of  
$10\;{\rm kpc}$, at the present time, as a result of the present modelling 
assumptions (see Figure 4). 

Other proposals for the amplification of the magnetic field in the center
of the cooling flow clusters are:  i) rotational driven mechanisms, in which the
twisting of the magnetic flux tubes and/or the operation of fast $\alpha-\omega$
dynamo are the responsible for the increase of the magnetic strength (Godon et
al.  1998); ii) turbulence induced amplification (Eilek 1990; Mathews \&
Brighenti 1997).  However, these processes can not account for the strong
magnetic fields observed in the center regions, confirming the expectations
previously discussed by authors like Goldshmidt \& Rephaeli (1993) and Carvalho
(1994).

\begin{figure}
\begin{center}
\vbox{
\psfig{file=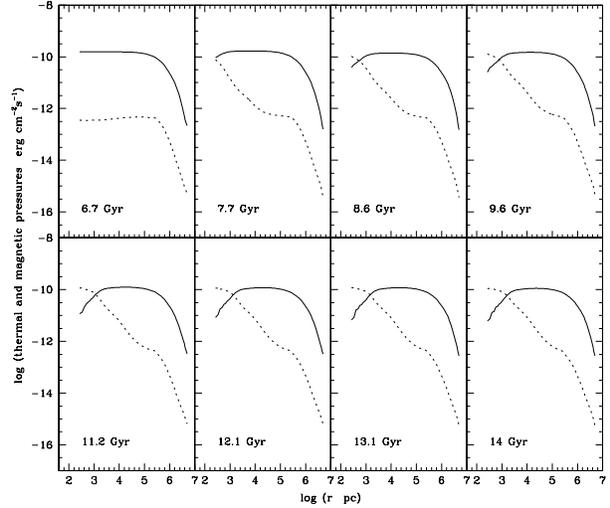,width=8.0truecm,bbllx=98pt,bblly=227pt,bburx=516pt,bbury=588pt}}
\end{center}
\caption[]{Evolution of the magnetic (dashed lines) and thermal (full lines)
pressure profiles on late stages of the ICM evolution for model I. 
Noting that the magnetic pressure increases until reaches
equipartition at inner regions of the cooling flow 
(at scales $\la 1\;{\rm kpc}$).}
\end{figure}

\begin{figure}
\begin{center}
\vbox{
\psfig{file=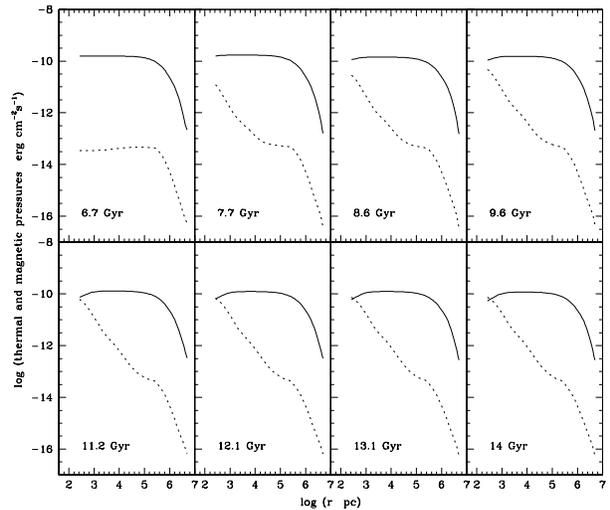,width=8.0truecm,bbllx=98pt,bblly=227pt,bburx=516pt,bbury=588pt}}
\end{center}
\caption[]{Evolution of the magnetic (dashed lines) and thermal (full lines)
pressure profiles on late stages of the ICM evolution for model II. 
Noting that the magnetic pressure increases until reaches
equipartition at inner regions of the cooling flow 
(at scales $\la 0.5\;{\rm kpc}$).}
\end{figure}

\section{Discussions and Conclusions}

The present models are in many aspects similar to the one of Soker \& Sarazin
(1990).  However there are two important differences between our model and
theirs:  i) they take into account only small-scale magnetic field effects; and
ii) they consider homogenous cooling flow.  Since we consider inhomogeneous
cooling flow (i.e.  $\dot{M}$ decreases with decreasing $r$) the amplification
of $B$ is smaller in our models.  As a matter of fact the magnetic pressure
reaches equipartition only at radius as small as $\ga 1\;{\rm kpc}$ (model I) or
$\ga 0.5\;{\rm kpc}$ (model II), because the central increase of the $\beta $
ratio is moderate in our model.  Our more realistic description of the field
geometry is crucial.  This implies that the effect of the magnetic pressure on
the total pressure of the intracluster medium, even on regions as inner as few
kpc, is small.  Tribble (1993) studying the formation of radio haloes in cooling
flow clusters from the point of view of the cluster evolution via mergers,
suggested typical magnetic field strengths of $\sim 1\;{\rm \mu G}$.  In
addition, Zoabi et al.  (1996), studying a completely different characteristic
of the ICM (magnetic fields on the support of X-rays clumps and filaments),
adopted the usually assumed magnetic to pressure ratio, at few scales of $%
10-20\;{\rm kpc}$, of $0.1$, and following a simple geometry of the field in
which it is amplified by the radial inflow, this ratio become $\sim 1$ at $%
\sim 5\;{\rm kpc}$.  Again our results are more or less compatible with the
above ones (for the cluster scale magnetic field), but the equipartition
condition is reached at smaller scales.

There are a number of papers discussing heating processes on the inner part
of the cooling flow clusters, in particular mechanisms to power the emission
lines of optical filaments, which use the magnetic energy transformed in
optical emission via magnetic reconnection (Jafelice \& Fria\c {c}a 1996) or
dissipation of Alfv\'{e}n waves (Fria\c {c}a et al. 1997). These works are
based in the enhancement of the magnetic pressure on scales smaller than $%
\sim 10\;{\rm kpc}$, where the filaments are observed (Heckman et al. 1989).
Finally, our results suggest that the effect of the
magnetic fields on the ICM dynamics can be relevant only on very small
scales: $\beta \sim 10^{-1}$, $ r\la 10 \;{\rm kpc}$, and $\beta \sim 1$, 
$r \la 1 \;{\rm kpc}$, depending on the model adopted (see Figure 7). 
From Figure 7 one can see quite clearly that the equipartition condition 
is reached at smaller radii for models in which $\beta_0$ is equal to $10^{-3}$ 
(model II and model IV), emphasizing the agreement between our models, another 
theoretical models, and observations. 

It is also quite relevant noting that the general agreement of our models 
and the available data can be emphasized by the fact that observations 
give us only limits on the magnetic field intensities. In the case of 
rotation measures the limit is the upper one, in contrast with the data 
coming from inverse Compton scattering which give the lower limit of this 
quantity. Therefore, from our best model (model I, see Figure 3) the expected 
field 
intensity is lower than the observed value (provided via rotation measures), 
on scales of $10\;{\rm kpc}$, and higher than the field intensity 
derived from X-ray inverse Compton scattering, on larger scales 
($700\;{\rm kpc}$). 

\begin{figure}
\begin{center}
\vbox{
\psfig{file=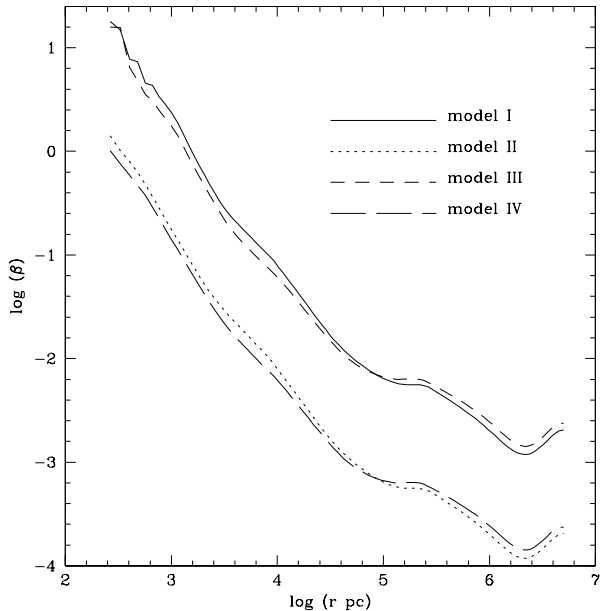,width=8.0truecm,bbllx=135pt,bblly=243pt,bburx=464pt,bbury=590pt}}
\end{center}
\caption[]{$\beta$ profiles on $14 \; {\rm Gyr}$ for models I - IV. The profiles 
are quite similars, except for the fact that models with $\beta_0 = 10^{-3}$ 
have final $\beta$-values lower. From the figure is also clear that the 
equipartition condition occurs at outer radii for higher $\beta_0$ models, and 
that anyway this condition is reached only on radii smaller than 
$\sim 1 \;{\rm kpc}$.}
\end{figure}

That the discrepancy found in the determination of mass, from gravitational
lensing and from X-rays observations (Loeb \& Mao, 1994; Loewenstein, 1994;
Miralda-Escud\'{e} \& Babul, 1995), and in the mass distribution of the
gravitating matter, mostly dark matter (Eyles et al.  1991), can be consequences
of the standard description of the ICM, in which it is assumed hydrostatic
equilibrium driven by thermal pressure (Fabian 1994), is a subject of
discussion.  Allen (1998) argued that, at least for cooling flow clusters, the
above discrepancy is resolved and, therefore, the effect of non-thermal
pressures on the hydrostatic equilibrium of these systems could be completed
discarded.  However, it is important to point out that the radius in which
magnetic pressure reaches equipartition is much smaller than the core or arc
radii obtained by Allen in his analysis ($\sim$ 50 kpc, in average), implying
that, despite Allen's results, at smaller scales the non-thermal pressures can
be important.

Theoretical models, like the one here presented, point
out that magnetic pressure does affect the hydrostatic equilibrium of the 
ICM, but only 
in the inner radius, as small as $\sim 1 \;{\rm kpc}$. 
In addition, it is important to remind that there are other sources of
non-thermal pressures that could be considered jointly to the magnetic
pressure before to close the discussion on whether or not non-thermal
pressures can explain the discrepancies on the mass estimations of the
galaxy clusters.

\subsection*{Acknowledgements}

One of the authors (D.R.G.) would like to thank the Brazilian agency FAPESP
(97/05246-3) for support, and the other author (A.C.S.F.) would like to
thank the Brazilian agency CNPq for partial support. We also would like to
acknowledge partial support by Pronex/FINEP (41.96.0908.00).

\end{document}